# The mechanism of recrystallization process in epitaxial GaN under dynamic stress field : Atomistic origin of planar defect formation

C. R. Das,[1] S. Dhara,[1,2,a)] H. C. Hsu,[3] L. C. Chen,[3] Y. R. Jeng,[4] A. K. Bhaduri,[1] Baldev Raj,[1] K. H. Chen,[3,5] S. K. Albert,[1]

[1]Metallurgy and Materials Group, Indira Gandhi Center for Atomic Research, Kalpakkam-603102, India

[2]Department of Electrical Engineering, Institute for Innovations and Advanced Studies, National Chen Kung University, Tainan-701, Taiwan.

[3] Center for Condensed Matter Sciences, National Taiwan University, Taipei-106, Taiwan

[4] Department of Mechanical Engineering, National Chung Cheng University, Chia-Yi 621, Taiwan

[5] Institute of Atomic and Molecular Sciences, Academia Sinica, Taipei-106, Taiwan

*Abstract*

The mechanism of recrystallization in epitaxial (0001) GaN film, introduced by indentation technique, is probed by lattice dynamic studies using Raman spectroscopy. The recrystallized region is identified by Micro-Raman area mapping. 'Pop-in' bursts in loading lines indicate nucleation of dislocations and climb of dislocations. These processes set in plastic motion of lattice atoms under stress field at the center of indentation for the initiation of recrystallization process. A planar defect migration mechanism is evolved. A pivotal role of vacancy migration is pointed out, for the first time, as the rate limiting factor for the dislocation dynamics initiating the recrystallization process in GaN.

**Keywords :** GaN, recrystallization, indentation, dislocation, Raman spectroscopy

[a)] Author to whom correspondence should be addressed; electronic mail: s_dhara2001@yahoo.com

## INTRODUCTION

GaN is one of the most important optoelectronic semiconductors in its applications as white light source, blue diode leading to UV lithography. The blue diode laser has an enormous potential for read and write head for high density digital versatile disk (DVD) memories and printing (typically calculated for 1200 dpi having ~17 μm spot size with 1 mm depth of field using 6 mm optics). Its most imminent application in white light based (using diode laser based RGB) compact display devices. However, high dislocation density in GaN is one of the major hindrances for its application as blue laser. Several techniques have been adopted for the diminution of dislocation densities in GaN,[1] including our recent report of recrystallization under indentation.[2] A layer-by-layer defect formation in GaN is studied using structural studies,[3] without detailed atomistic model for its origin and dynamics. It necessitates a detailed study of lattice dynamics in the stress free region, as well as in the strained volume for comparison. Investigation of phonon modes in Raman scattering process is well established, and is one of the potential methodologies to understand lattice dynamics at the atomic level.

We report here the mechanism of recrystallization in wurtzite (WZ) epi-GaN film under indentation. Lattice dynamic studies by probing phonon modes reveal the possible mechanism of defect migration, which actually influences the recrystallization process. Defect nucleation and its dynamics in GaN are also studied for an atomistic view of the process where role of both point (vacancy) and extended (dislocation) defects is noted for the first time.

## EXPERIMENTAL

An undoped epi-layer of 6 μm thickness (0001) GaN/$Al_2O_3$ grown by MOCVD with threading dislocation (TD) <5x10$^8$ cm$^{-2}$ (TDI, USA) along with intrinsic carrier concentration of ~3x10$^{17}$ cm$^{-3}$ is used for the present study. The sample is indented using the micro-indenter with a Berkovich diamond tip. The data presented in this report exploits indentation conditions of load 100-400 mN; same loading-unloading rate of 1-50 mN.s$^{-1}$ and holding time 5 s. Excitation wavelength of 632.8 nm of He-Ne laser is used for the micro-Raman spectroscopic studies (probed volume < 1 µm$^3$) in the back scattering geometry. A liquid nitrogen cooled CCD detector is used for recoding the scattering intensity.

## RESULTS AND DISCUSSION

### LATTICE DYNAMICS OF GaN UNDER INDENTATION

The structural transformation is studied close to the indented region using micro-Raman spectroscopy (Fig. 1a). The $E_2$ (high) mode at 570 cm$^{-1}$, measured outside the indented region, resembles the reported value of epi-GaN on sapphire substrate. The inset shows the spots measured out- and insides of the indentation mark recorded in the optical microscope attached to the spectrometer. However, micro-Raman measurements inside the indented region show redshift of phonon mode to 567 cm$^{-1}$ gradually from interface region (edge of the spot) to the center of indented spot. This value is close to the calculated and measured value for $E_2$ (high) phonon of bulk GaN and GaN nanostructure under stress free conditions.[4,5] Double peaks are observed for the interface region close to the edge of the indentation (Fig. 1a), showing contributions from both the stressed region outside and stress free region inside the indented region. Raman area mapping (outset of Fig. 1a) using spectral part of 568-573 cm$^{-1}$ shows red (bright in the grayscale) region lying outside the indented region and 562-568 cm$^{-1}$ shows green (bright in the grayscale) region lying inside the indented region. It clearly shows that the 567 cm$^{-1}$ signal originates from the stress free region inside the indented mark and the 570 cm$^{-1}$ peak originates from the stressed region of the sample.

Two additional peaks at 531 cm$^{-1}$ and 559 cm$^{-1}$ (Fig. 1a), corresponding to $A_1$(TO) and $E_1$(TO) modes respectively, are also observed in the indented region. According to the selection rule in WZ crystal of GaN, TO phonon modes are forbidden in the backscattering geometry for the (0001) oriented planes.[5] However, small disorientations of crystallites in the indented region allow the phonon modes corresponding to other crystalline orientations, as shown in the schematics (Fig. 1b) for the WZ crystal.[6,7] It can be also mentioned that the phonons (Fig. 1b) corresponding to the $E_2$(high), and TO modes belonging to $A_1$ and $E_1$ symmetries are along $X$ and/or $Y$ direction (confined mostly to the $XY$ plane). Therefore, the lattice modes in the $XY$ plane are vibrant. After indentation, the surface of the epi-GaN is no longer flat and in the 'V' grooved geometry a right angled scattering is not obscured. Therefore, we have considered both the backward and right angled scattering processes. From the evolution of $E_2$ (high) and TO phonon modes, collected in the backscattering geometry, it seems that stress is released in the planar direction so that Raman modes along $X$ or $Y$ or both gets modified. Interestingly, the peak position ~ 736 cm$^{-1}$ corresponding to $A_1$(LO) mode, in the different regions close to the indentation spot (Fig. 2a), shows no shift in the position. The Raman scattering configuration in the WZ crystal (Fig. 2b),[6,7] for a phonon corresponding to LO mode of $A_1$ symmetry is always (leaving quasi LO modes in the right angled scattering process) along $Z$ direction (normal to $XY$ plane). Thus with no change in the peak position of $A_1$ (LO) mode, it is obvious that lattice modes normal to $XY$ plane remain unaltered. It is also well known that defects in GaN propagate in the planar direction. A layer-by-layer model of defect accumulation in epi-GaN is envisaged from Rutherford backscattering (RBS) based channeling experiments supported by cross-sectional transmission electron microscopic imaging.[3] Thus, the overall evolution of phonon modes may be due to nucleation of dislocation, and release of its stress field under indentation stress to set in planar motion at the centre of indentation region by dislocation climb. Dislocation climb in the material are reported under very high hydrostatic stress due to indentation, even though test temperature is a relatively small fraction of the melting temperature of material.[8]

A small peak ~707 cm$^{-1}$ peeps in the spectra collected from the regions inside the indentation. Changes are observed in lattice vibration with the size of the crystal reducing to few nanometers. It is prominent where the phonon is confined to the surface, giving rise to a wavenumber that falls between TO and LO modes.[9] The dielectric constant ε(ω) is negative in between the LO and TO mode. This is the region where the surface optic (SO) phonon appears and the electromagnetic field associated with the SO phonon is localized at the surface of the material. The surface phonon frequency and intensity mainly depend on the size and the shape of the nanostructured material. The expression for SO phonon in spherical clusters is given by,[9]

$$\omega^2_{so} = \omega^2_{TO} \frac{\varepsilon_0 + 2\varepsilon_m}{\varepsilon_\infty + 2\varepsilon_m} \quad \text{------------------------------------- (1)}$$

Where $\varpi_{TO}$ is the frequency of TO phonon, $\varepsilon_0$ and $\varepsilon_\infty$ are the static and high frequency dielectric constant of the material and $\varepsilon_m$ is the dielectric constant of the medium. Here, we use 10.4 and 5.8 for the values of $\varepsilon_0$ and $\varepsilon_\infty$ respectively for GaN.[10] The dielectric constant of the medium (air) is taken as 1. Taking $A_1$(TO) at 561 cm$^{-1}$ for GaN in Equation 1 we get SO phonon wavenumber at 707 cm$^{-1}$, which matches the observed peak (Fig. 2a). Thus the origin of this new peak can be assigned to the SO mode of GaN.

**NUCLEATION AND DYNAMICS OF DEFECTS UNDER INDENTATION STRESS FIELD**

Further, we explore the role of nucleation as well as the motion of dislocation and rate limiting factors in the recrystallization process. A detailed description of compliance curve along with 'pop-in' bursts in the loading line is reported in our earlier studies,[2] and also shown in supplementary Fig. S1 for high and low loads. Recent findings suggest that the origin of an initial 'pop-in' in crystals could be explained in terms of homogeneous,[11-13] or heterogeneous[14] dislocation nucleation under the penetrating tip. High values of hydrostatic and shear stress are present within the plastic volume beneath the indentation and the later is responsible for the plasticity. After the indenter pressed into the material, the instant plasticity ('pop-in') occurs as soon as shear stress crossed the theoretical stress (estimated in supplementary Appendix A for the present study). The mechanism responsible for the 'pop-in' bursts appear to be associated with the nucleation and movement of dislocation sources.[14,15]

A hardness value of ~10 GPa is achieved in our indentation study,[2] and the value is comparable to bulk value of GaN.[16] It is also observed that the hardness of GaN film is also sensitive to the loading rate (strain rate) as shown in Fig 3(a). At a lower strain rate, hardness is found to be ~10 GPa and increases with strain rate. Increasing hardness with strain rate can be explained with increase in dislocation density due to its multiplication[17] by shear stress component following strain gradient plasticity model.[18] Stress relaxation at the tip of the indentation can be understood from the strain rate exponent,[19] $m = \partial \ln H / \partial \ln \dot{\varepsilon}$) and activation volume,[13,18] $V_A = 3\sqrt{3k_B T}(\partial \ln \dot{\varepsilon}/ \partial H)$ ( Boltzmann constant, $k_B$ and absolute temperature, $T$) analysis. These two parameters are used to determine possible deformation mechanism for a given material. It also provides quantitative measures of the sensitivity of hardness, $H$ to the strain rate, $\dot{\varepsilon}$ and insight to the deformation mechanisms. The values of $m$ and $V_A$ are calculated to be around 0.26 and $0.3b^3$ from the linear plots of $\ln H$ vs. $\ln \dot{\varepsilon}$ (Fig. 3b) and $\ln \dot{\varepsilon}$ vs. $H$ (Fig. 3c), respectively, at the peak load of 100 mN. The value of $V_A$ is comparable to the average vacancy volume ($\sim 0.7b^3$) in GaN considering both Ga and N vacancy are present.[20] It can be mentioned that with increase in peak load of 200 mN, the value of $m$ decreases to 0.06 [supplementary Fig. S2(a)] and $V_A$ are found to increases to $1.6b^3$ [supplementary Fig. S2(b)]. It is generally accepted that high $m$ is indicative of smaller $V_A$.[21] Thus, the magnitude of $V_A$ extracted from our experiment are reflective of atomic-scale event; therefore, one can think of point defect (vacancies)- related process is the rate limiter for the plastic deformation process.

Atomistic simulation studies on perfect single crystals have shown that dislocation nucleate either homogeneously,[11-13] or heterogeneously[14] beneath the indenter in the material. $V_A$ ~ $0.5b^3$ is reported for homogeneous dislocation nucleation in single crystal Pt.[22] Though self-diffusion is slow at ambient temperature, authors argue that rise in temperature for the material beneath the indenter due to adiabatic process,[23] and high pressure gradient can cause the high diffusivity of vacancy to the high pressure region to release the compressive stress from the material. In a separate study, the same idea is also proposed by Schuh *et al.*[14] The vacancy concentration (considering intrinsic carrier primarily contributed from vacancy like defects in GaN) can be estimated ~$10^{17}$ cm$^{-3}$. Lower activation volume and high vacancy concentration, from the pre-existing defects such as vacancies, suggest heterogeneous dislocation nucleation model. The formation of homogenous dislocation loop may not be possible due to large requirement of activation energy in this process.[14,22] A stress-induced crystallization has been also reported in Ge at temperature as low as 400 K.[24] Observation of in-plane phonon dynamics and low activation volume for vacancy migration suggests that climb of dislocation by vacancy is the rate limiting factor for stabilizing the recrystallization process.

**CONCLUSION**

Here, we address an important issue of recrystallization process in GaN under indentation stress. Heterogeneous nucleation of dislocation, originated by shear stress within the plastic zone, is evidenced in the indentation process. The release of stress takes place by climb of dislocation due to high hydrostatic stress present in the plastic volume. Vacancy migration is found to equilibrate the recrystallization process by limiting the dislocation motion. The mechanism of the crystallization process is clearly evaluated with lattice dynamics studies. The present study implies important clues for reducing residual stress in a GaN epi-film, and the findings will have broad technological repercussion.

ACKNOWLEDGEMENT

One of the authors (CRD) would like to thank C. Phaniraj of IGCAR, India for his valuable discussion.

**Figure captions**

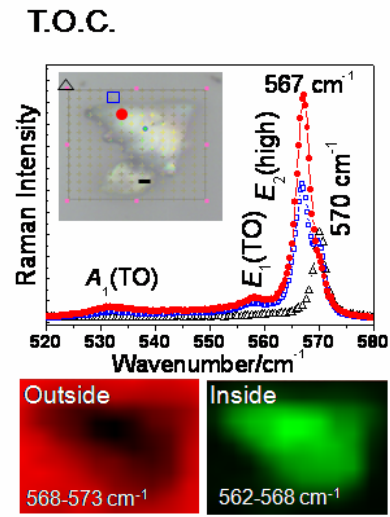

**TOC :** The mechanism of recrystallization in epitaxial GaN film, introduced by indentation technique, is probed by lattice dynamic studies using Raman spectroscopy. 'Pop-in' bursts in loading lines indicate nucleation of dislocations and climb of dislocations, setting in plastic motion of lattice atoms under stress field for the initiation of recrystallization process. A planar defect migration mechanism is evolved with vacancy migration is considered as the rate limiting factor for the dislocation dynamics initiating the recrystallization process in GaN.

Fig. 1

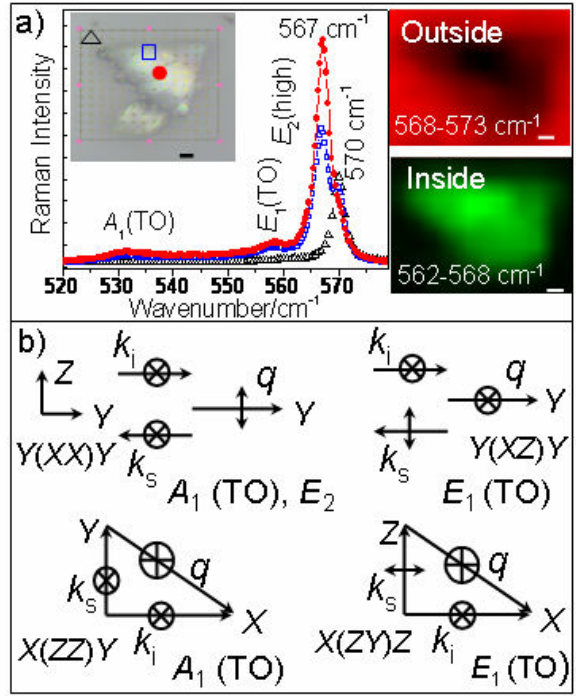

**Figure 1**. (Color online) a) Micro-Raman spectra for epi-GaN outside and different regions inside the indentation spot. Inset shows corresponding optical image of the indentation spot. Area mappings of outside and inside of the indentation spot, using different spectral regions indicated in the picture, are shown at the outset. Scale bar is 1 μm. b) Different Raman scattering configuration for WZ crystal in the backward (top) and right angled (botttom) direction corresponding to $E_2$, and TO modes of $E_1$ and $A_1$ symmetries for usual incident and scattering notations. (*Drawn after* Figure 6 of Ref. [6] *permission taken from American Physical Society*)

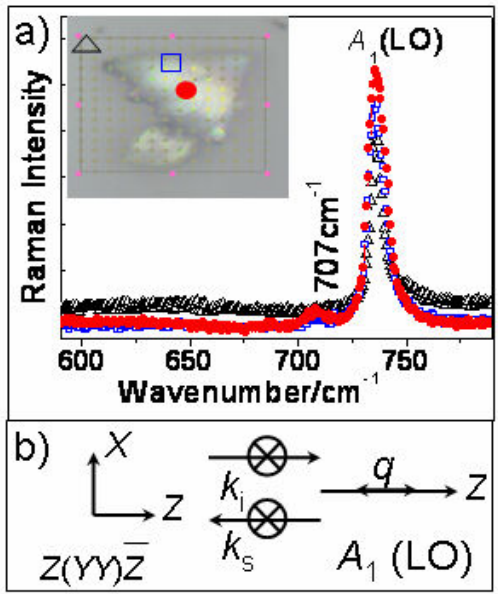

**Figure 2**. (Color online) a) Micro-Raman spectra for epi-GaN outside and different regions inside the indentation spot. Inset shows corresponding optical image of the indentation spot. Scale bar is 1 μm. b) Raman scattering configuration for WZ crystal in the backward scattering direction corresponding to LO mode of $A_1$ symmetry for usual incident and scattering notations. (*Drawn after* Figure 6 of Ref. [6] *permission taken from American Physical Society*)

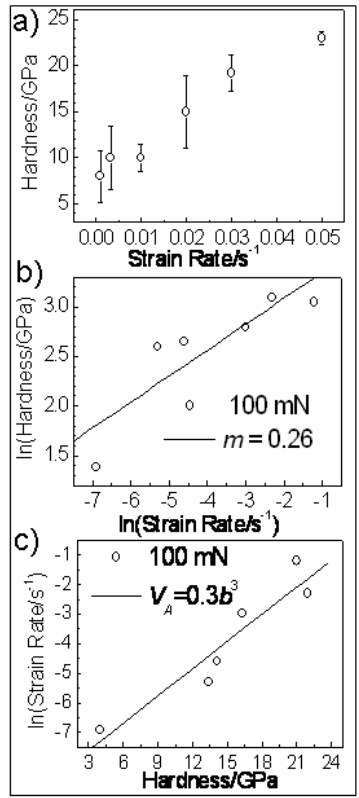

**Figure 3**. (a) Variation of hardness with indentation strain rate. Typical (b) log-log plot of hardness vs strain rate, and (c) log (indentation strain rate) vs hardness plot for the peak load of 100mN

**Supplementary :**

*Appendix A: Calculation for shear stress at critical load*

The maximum shear stress ($\tau_{max}$) occurring in the film at critical load, $P_{criti}$ can be estimated by [Thokala R, Chaudhuri J. *Thin Solid Films* 1995; **266**: 189],

$$\tau_{max} = 0.12 \left( \frac{P_{criti} E_s^2}{R^2} \right)^{1/3} \quad \text{-----------------------} (1)$$

Where, $R$ is the tip radius of curvature of the indenter, $\tau_{max}$ is the shear stress, $P_{criti}$ is the initial bursting load, $E_s$ is the modulus of the material can be estimated from

$$\frac{1}{E_R} = \left( \frac{(1-\upsilon_{in}^2)}{E_{in}} + \frac{(1-\upsilon_s^2)}{E_s} \right) \quad \text{--------------------------}(2)$$

where $E_R$ is reduced modulus, $E_{in}$ (1141GPa) is indenter modulus and $\nu_{in}$ (=0.07) is indenter Poisson ratio, $\nu_s$ is material poison ratio is 0.352. The observed value for $P_{criti}$ = 10.69 mN (Supplementary Fig. S1b) for one of the studies with very low load showing distinct threshold 'pop-in'. $\tau_{max}$ can be estimated to be ~ 12.42 GPa from Equation 2 with $E_s$ value of 198 GPa, as calculated from Equation 2 and $R = $ ~ 600 nm (measured from the SEM image). This value is higher than the calculated theoretical shear strength $\tau_{th}$ [= $\mu_s/2\pi$, where $\mu_s$ =$E_s/2(1+\nu_s)$= 73 GPa], value ≈11.66 GPa to initiate 'pop-in'.

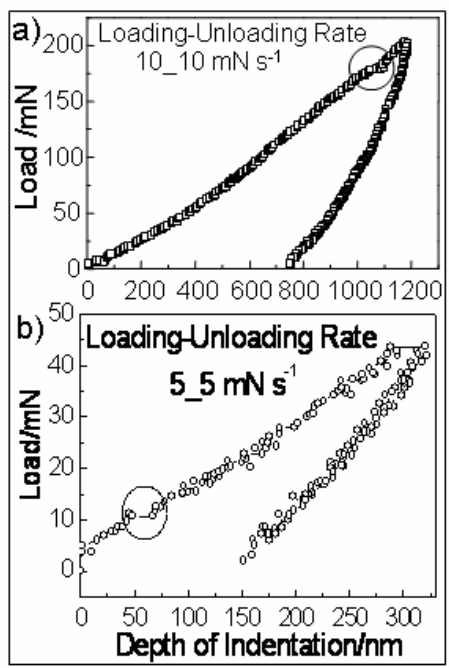

**Figure S1**. Variation of load and depth of indentation at peak loads of (a) 200 mN (high) and (b) 45 mN (low)

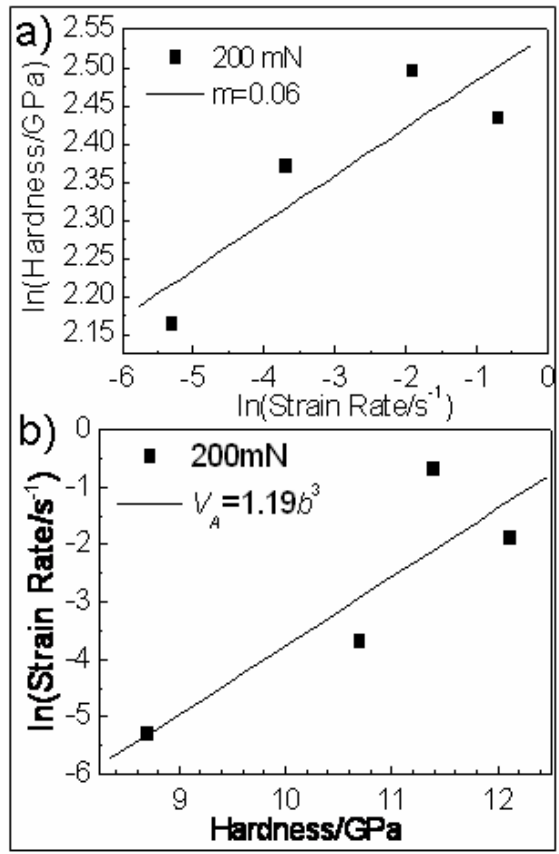

**Figure S2.** Typical (a) log-log plot of hardness vs strain rate, and (b) log (indentation strain rate) vs hardness plot for the peak load of 200mN